\documentclass[%
 reprint,
superscriptaddress,
 amsmath,amssymb,
prb,
]{revtex4-1}

\usepackage{dsfont}
\usepackage{graphicx}
\usepackage{hyperref}
\usepackage{color}

\usepackage{mathrsfs}

\usepackage{braket}

\usepackage{overpic}
\usepackage{tikz}
\usetikzlibrary{decorations.pathreplacing}

\newcommand{\im}{\mathrm{i}}

\definecolor{mygreen}{rgb}{0,0.5,0}
\definecolor{myblue}{rgb}{0,0,0.75}
\definecolor{mymagenta}{cmyk}{0,1,0,0.12}

\usepackage{umoline}

\begin{document}
\title{Semiclassical echo dynamics in the Sachdev-Ye-Kitaev model}

\author{Markus Schmitt}
\email{markus.schmitt@berkeley.edu}
\affiliation{%
 Institute for Theoretical Physics, 
	Georg-August-Universit\"at G\"ottingen, 
	Friedrich-Hund-Platz 1, 37077 G\"ottingen, Germany
}
\affiliation{%
 Max Planck Institute for the Physics of Complex Systems, 
	N\"othnitzer Stra\ss e 38, 01187 Dresden, Germany
}
\affiliation{%
 Department of Physics, University of California, Berkeley, CA 94720, USA
}
\author{Dries Sels}
\affiliation{%
 Department of Physics, Boston University, 590 Commonwealth Ave., Boston, MA 02215, USA
}
\affiliation{
Harvard University, 17 Oxford Street, Cambridge, MA 02138, USA
}
\affiliation{
Theory of quantum and complex system, Universiteit Antwerpen, Universiteitsplein 1, Antwerpen, BE
}
\author{Stefan Kehrein}
\affiliation{%
 Institute for Theoretical Physics, 
	Georg-August-Universit\"at G\"ottingen, 
	Friedrich-Hund-Platz 1, 37077 G\"ottingen, Germany
}
\author{Anatoli Polkovnikov}
\affiliation{%
 Department of Physics, Boston University, 590 Commonwealth Ave., Boston, MA 02215, USA
}
\date{\today}

\begin{abstract}
The existence of a quantum butterfly effect in the form of exponential sensitivity to small perturbations has been under debate for a long time. Lately, this question gained increased interest due to the proposal to probe chaotic dynamics and scrambling using out-of-time-order correlators. In this work we study echo dynamics in the Sachdev-Ye-Kitaev model under effective time reversal in a semiclassical approach using the truncated Wigner approximation, which accounts for non-vanishing quantum fluctuations that are essential for the dynamics. We demonstrate that small imperfections introduced in the time-reversal procedure result in an exponential divergence from the perfect echo, which allows to identify a Lyapunov exponent $\lambda_L$. In particular, we find that $\lambda_L$ is twice the Lyapunov exponent of the semiclassical equations of motion. This behavior is attributed to the growth of an out-of-time-order double commutator that resembles an out-of-time-order correlator.
\end{abstract}

\maketitle

\section{Introduction}
The question of chaos and the possibility of a butterfly effect 
in quantum systems is a long-standing problem that received increased
attention in recent years.
In studies addressing the information paradox of black holes so-called
out-of-time-order correlators (OTOCs) of the form
\begin{align}
	\braket{\hat V(0)^\dagger\hat W(t)^\dagger\hat V(0)\hat W(t)}_\beta
	\label{eq:otoc}
\end{align}
were introduced to probe the sensitivity of the dynamics to small perturbations and
scrambling, i.e., the delocalization of initially local information \cite{Hayden2007,Sekino2008,Shenker2014,Kitaev2014}.
A semiclassical analysis of the OTOC motivates that it can exhibit exponential growth,
allowing to identify a Lyapunov exponent \cite{Larkin1969}.
In fact, it was found that in a black hole theory OTOCs grow exponentially with the maximal
possible rate $\lambda_L=\frac{2\pi}{\beta}$. \cite{Maldacena2016}
Remarkably, there exists a solvable model of interacting fermions, which also saturates this
bound at low temperatures, namely the Sachdev-Ye-Kitaev (SYK) model \cite{Kitaev2015,Maldacena2016a}, which is a variant of a model originally introduced by Sachdev and Ye \cite{Sachdev1993,Sachdev2010}.

OTOCs as dynamical probe of chaos and scrambling are also of interest in condensed matter
systems beyond the AdS/CFT paradigm \cite{Bohrdt2017,Huang2017,Chen2017,Iyoda2017,Swingle2017,Shen2017,Patel2017,Fan2017}.
Particularly intriguing is the connection to the question how and in what sense closed quantum 
many-body systems thermalize when
initially prepared far from equilibrium, which has been studied with great efforts in recent years \cite{DAlessio2016,Gogolin2016}.
The corresponding statistical description of the stationary state is only justified if the information
about the initial state cannot be recovered in practice, i.e., the dynamics is irreversible.

To assess the irreversibility of the dynamics a common approach is to study imperfect effective time
reversal.
In classical systems it is understood that the exponential sensitivity of the dynamics to small
perturbations prohibits recovery of the initial state, because perfect time reversal is impossible in practice
\cite{Thompson1874,Boltzmann1872,Loschmidt1876,Boltzmann1877}.
Under chaotic dynamics any imperfection occurring in the time reversal operation leads to an exponential divergence from accurate recovery of the initial state with a rate that is largely independent of the perturbation,  namely the Lyapunov exponent.
This renders the improvement of the protocol prohibitively expensive.

Analogous approaches have been explored considering quantum systems.
In few-body systems the decay characteristics of the Loschmidt echo $\mathcal L(\tau)=|\braket{\psi_0|\hat U_E^\epsilon(\tau)|\psi_0}|^2$ with the echo operator $\hat U_E^\epsilon(\tau)=e^{\im(\hat H+\epsilon\hat V)\tau}e^{-\im\hat H\tau}$, where $\epsilon\hat V$ constitutes a small perturbation to the Hamiltonian, were used as indicator of chaos and irreversibility \cite{Peres1984,Gorin2006,Jacquod2009}.
For many-body systems, however, overlaps lack experimental significance.
Instead, the decay of observable echos under imperfect effective time reversal was studied to investigate irreversibility \cite{Fine2014,Elsayed2015,Zangara2015,Schmitt2016,Schmitt2017}.

In the works mentioned above the focus was on decay laws occurring in the echo dynamics at late times.
By contrast, imperfect effective time reversal in classical systems features initial dynamics that is governed by the butterfly effect.
The possibility of a butterfly effect that occurs analogously in quantum systems is currently under debate \cite{Fine2014,Fine2017,Rozenbaum2017,Bohrdt2017,Scaffidi2017,Patel2017,Khemani2018,Swingle2018}.
Moreover, the realization of effective time reversal was recently reported from an experiment with trapped ions, where OTOCs were measured in the form of echo dynamics \cite{Gaerttner2017}.

In this work we study the dynamics of the SYK model under imperfect effective time reversal in
a semiclassical approach using the truncated Wigner approximation.
We demonstrate that the small imperfection leads to an exponential divergence from the perfect echo.
This divergence can be attributed to the exponential growth of an out-of-time-order double commutator
similar to an OTOC and it allows to identify a Lyapunov exponent based on the echo dynamics.

The structure of the paper is as follows: First, in Sections \ref{sec:time_reversal} and \ref{sec:hamiltonian} we introduce the echo protocol under consideration and the model of interest. Section \ref{sec:semiclassical_dynamics} comprises an introduction to the truncated Wigner approximation and a discussion of its applicability to the SYK model. In Section \ref{section:echoes} we present our results for the echo dynamics in the semiclassical limit. Before the final discussion in Section \ref{sec:discussion} we include in Section \ref{sec:TWAvsMF} an extended elaboration on the distinction between mean field dynamics and the truncated Wigner approximation in the context of the SYK model.

\section{Imperfect effective time reversal}
\label{sec:time_reversal}
In the following we will investigate the echo dynamics of an observable $\hat O$ under imperfect effective time reversal.
The perturbation is introduced by the action of a perturbation operator $\hat P_\epsilon$ on the time-evolved state at the point of time reversal.
Here $\epsilon$ denotes a parameter for the smallness of the perturbation.
A natural choice for the perturbation operator is unitary time evolution for a short interval $\delta t$ with a perturbation Hamiltonian $\hat H_p$, i.e., $\hat P_{\delta t}=e^{-\im\hat H_p\delta t}$.
The quantity of interest is the echo signal
\begin{align}
	E_{\hat O}(\tau)=\braket{\psi_0|\hat U_E^{\delta t}(\tau)^\dagger \hat O\hat U_E^{\delta t}(\tau)|\psi_0}
	\label{eq:echo_definition}
\end{align}
with the echo operator $\hat U_E^{\delta t}(\tau)=e^{\im\hat H\tau}\hat P_{\delta t}e^{-\im\hat H\tau}$.
This constitutes an OTOC in the case that the initial state is an eigenstate of the observable, $(\hat O-\mu)\ket{\psi_0}=0$ \cite{Gaerttner2017,Schmitt2017}.
Moreover, expanding the echo operator $\hat U_E^{\delta t}(\tau)$ in orders of $\delta t$ yields
\begin{align}
	\Delta &E_{\hat O}(\tau)=\braket{\psi_0|\hat O|\psi_0}-E_{\hat O}(\tau)\nonumber\\
	&=\im\delta t\braket{\psi_0|[\hat H_p(\tau),\hat O]|\psi_0}\nonumber\\
	&\quad
	+\frac{\delta t^2}{2}\braket{\psi_0|[\hat H_p(\tau),[\hat H_p(\tau),\hat O]]|\psi_0}
	+\mathcal O(\delta t^3)
	\label{eq:echo_divergence}
\end{align}
for the divergence from the perfect echo.
In this expression the linear term corresponds to linear response and it vanishes in the case that the initial state is an eigenstate of the observable.
Hence, the quadratic term constitutes the leading contribution to the divergence from the perfect echo, accounting for the sensitivity of the dynamics to small perturbations.
Using the example of the SYK Hamiltonian
we will demonstrate in the following that this double commutator in fact determines the initial decay of the echo and that the corresponding divergence grows exponentially in time, which allows to identify a Lyapunov exponent.

\section{Model Hamiltonian}
\label{sec:hamiltonian}
The Hamiltonian of the fermionic SYK model is given by
\begin{align}
	\hat H=\frac{1}{(2N)^{3/2}}\sum_{ijkl}J_{ij;kl}\hat c_i^\dagger\hat c_j^\dagger\hat c_k\hat c_l\ ,
\end{align}
where the $J_{ij;kl}$ are complex-valued Gaussian random couplings with vanishing mean and variance $\sigma^2=\overline{|J_{ij;kl}|^2}$. 
$N$ denotes the number of fermionic modes.
The SYK model has a number of remarkable properties. 
Although strongly interacting, it is exactly solvable in the limit of large $N$.
At low temperatures it exhibits an emergent conformal symmetry indicating the existence of a
holographic dual \cite{Maldacena2016a}.
In this regime it is maximally chaotic in the sense that the Lyapunov exponent occurring in OTOCs saturates the bound that was derived for AdS black holes \cite{Maldacena2016}.


\section{Semiclassical dynamics in the SYK model}
\label{sec:semiclassical_dynamics}
We will analyze echo dynamics using the fermionic version of
the truncated Wigner approximation (TWA), which was recently developed in Refs.\ \cite{Davidson2017,Davidson2017a}.

\subsection{On the applicability of the truncated Wigner approximation}
\label{sec:onTWA}
The TWA is the saddle point approximation for the Keldysh action describing the Heisenberg evolution of the observables. It can be generally derived using standard path integral methods.\cite{Polkovnikov2010} Within the TWA time evolution of phase space variables is governed by the classical Hamiltonian equations of motion, which have to be supplemented by fluctuating initial conditions. In turn those are encoded in the Wigner function describing the initial state. Within the accuracy of the TWA one can generally approximate this Wigner function by a Gaussian capturing means and fluctuations of the phase space variables. Note that while formally classical equations of motion are identical to the Dirac mean field equations of motion (see Section \ref{sec:TWAvsMF}), the TWA reduces to the mean field approximation only if fluctuations in initial conditions are asymptotically vanishing with the saddle point parameter. This is, e.g., the case for initial coherent states or for polarized quantum spins in the large S-limit. But it is not the case, e.g., for stationary states of a high energy particle in a confining potential where the Wigner function approaches the broad in space micro canonical distribution rather than a single phase space point. In many instances, in particular when we deal with Fermions or spin one half degrees of freedom fluctuations are always large such that the mean field approximation is generally incorrect and moreover is not approached as the saddle point parameter increases (see, e.g., Refs.\ \cite{Altland2009,Orioli2017}). In the SYK model the large N limit ensures validity of the saddle point approximation\cite{Maldacena2016a,Eberlein2017} and, therefore, it is natural that the fermionic version of TWA will be asymptotically exact in the large N limit, which as we show in Section \ref{sec:accuracy} is indeed the case. Hence, $N$ serves as effective $\hbar^{-1}$.

\subsection{Phase space approach for Fermions}\label{subsec:eom}
Within the fermionic TWA a phase space representation is constructed for the fermionic bilinears, which satisfy the commutation relations of $so(2N)$ \cite{Davidson2017,Davidson2017a}; see also Ref. \cite{Yaffe1982} for a general picture of classical representations of quantum models.
The Weyl symbols of the fermionic bilinears are
$\tau_{\alpha\beta}=\big(\hat c_\alpha\hat c_\beta\big)_W=-\big(\hat c_\alpha^\dagger\hat c_\beta^\dagger\big)_W^*$
and
$\rho_{\alpha\beta}=\frac12\big(\hat c_\alpha^\dagger\hat c_\beta-\hat c_\beta\hat c_\alpha^\dagger\big)_W$.
The corresponding Weyl symbol of the SYK Hamiltonian expressed in terms of pairing operators is
\begin{align}
	\mathcal H&=\frac{1}{(2N)^{3/2}}\sum_{ijkl}J_{ij;kl}
	\Big(\tau_{ji}^*\tau_{kl}
	+\rho_{jk}\delta_{il}+\rho_{il}\delta_{kj}\Big)\ .
\end{align}

Generally, for phase space variables $X_\alpha$ of operators $\hat X_\alpha$, which obey some algebra
\begin{align}
	[\hat X_\alpha,\hat X_\beta]=\im f_{\alpha\beta\gamma}\hat X_\gamma
\end{align}
with structure constants $f_{\alpha\beta\gamma}$, the classical equations of motion are determined by
\begin{align}
	\frac{dX_\alpha}{dt}=f_{\alpha\beta\gamma}\frac{\partial (\hat H)_W}{\partial X_\beta}X_\gamma\ ,
\end{align}
where $(\hat H)_W\equiv\mathcal H$ is the Weyl symbol of the Hamiltonian.\cite{Davidson2017}

For the phase space variables of the fermionic bilinears and the Hamiltonian of the SYK model this yields
\begin{widetext}
\begin{align}
	\im\frac{d}{dt}\rho_{\alpha\beta}&=
	\Bigg(
	-\frac{\partial\mathcal H}{\partial\rho_{\gamma\alpha}}\rho_{\gamma\beta}
	+\frac{\partial\mathcal H}{\partial\tau_{\gamma\alpha}}\tau_{\beta\gamma}
	-\frac{\partial\mathcal H}{\partial\tau_{\alpha\gamma}}\tau_{\beta\gamma}
	\Bigg)
	-\Bigg(\alpha\leftrightarrow\beta\Bigg)^*\ ,
	\nonumber\\
	&=
	\frac{2}{N^{3/2}}\sum_{ijkl}J_{ijkl}\delta_{\alpha l}\Big(\tau_{ji}^*\tau_{\beta k}+\delta_{ik}\rho_{j\beta}\Big)-(\alpha\leftrightarrow\beta)^*
	\nonumber\\
	\im\frac{d}{dt}\tau_{\alpha\beta}&=
	\Bigg(
	\frac{\partial\mathcal H}{\partial\rho_{\alpha\gamma}}\tau_{\gamma\beta}
	+\frac{\partial\mathcal H}{\partial\tau_{\gamma\alpha}^*}\rho_{\gamma\beta}
	-\frac{\partial\mathcal H}{\partial\tau_{\alpha\gamma}^*}\tau_{\gamma\beta}
	\Bigg)
	-\Bigg(\alpha\leftrightarrow\beta\Bigg)
	\nonumber\\
	&=
	\frac{2}{N^{3/2}}\sum_{ijkl}J_{ijkl}\delta_{\alpha j}\Big(\delta_{il}\tau_{k\beta}-\tau_{kl}\rho_{i\beta}\Big)-(\alpha\leftrightarrow\beta)
	\ .
	\label{eq:twa_eom}
\end{align}
\end{widetext}

In the following we will consider uncorrelated initial states that are fully characterized by orbital
occupation numbers $n_\alpha=\braket{\hat c_\alpha^\dagger\hat c_\alpha}$.
In that case the Wigner function is well approximated by a multivariate Gaussian fixed by the first and second moments \citep{Davidson2017a}.
We will be interested in the expansion dynamics starting from an initially imbalanced occupation similar to  the situations studied in different recent cold atom experiments \cite{Schneider2012,Choi2016,Bordia2017}.
Given this initial state a suited observable to consider in the view of echo dynamics is the occupation imbalance
\begin{align}
	\hat M=\frac{1}{N}\sum_{\alpha=1}^N(2n_{\alpha}^0-1)(2\hat c_\alpha^\dagger\hat c_\alpha-1)
\end{align}
with Weyl symbol $\mathcal M=\frac{2}{N}\sum_{\alpha=1}^N(2n_\alpha^0-1)\rho_{\alpha\alpha}$, where $n_\alpha^0$ is the initial value of $n_\alpha$.

\subsection{Accuracy of the TWA}
\begin{figure}[ht]
\includegraphics[width=.9\columnwidth]{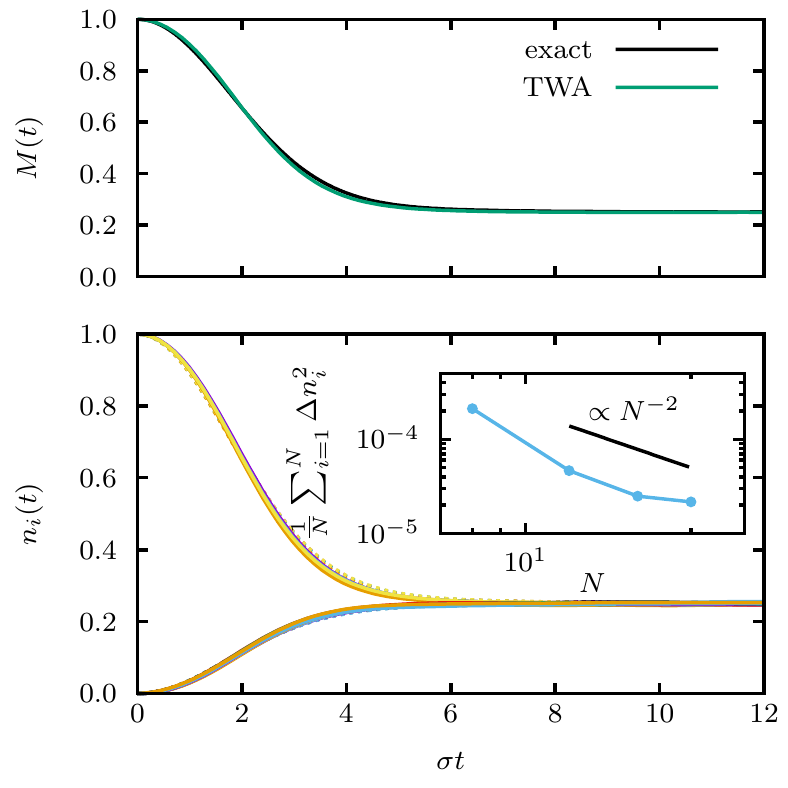}
\caption{Comparison of TWA results to the exact dynamics. The top panel shows the time evolution of the occupation imbalance with $N=20$ modes, whereas in the bottom panel the individual mode occupation numbers are shown. In the bottom panel dashed lines correspond to the exact result. The inset shows the system size dependence of the time-averaged squared deviation of the TWA from the exact result.}
\label{fig:twa_precision}
\end{figure}
\label{sec:accuracy}
In order to assess the accuracy of the TWA we compare the result for expansion dynamics from an uncorrelated initial state, where one quarter of the modes is occupied and the rest is empty, with exact dynamics.
Fig.\ \ref{fig:twa_precision} displays the corresponding time evolution of the occupation imbalance $M(t)$ and the individual mode occupations $n_i(t)=\braket{\psi(t)|\hat c_i^\dagger\hat c_i|\psi(t)}$ for $N=20$ and a disorder average involving 20 realizations.
The dynamics computed using TWA is in good agreement with the exact dynamics.
We find empirically that the accuracy of TWA improves as $N$ is increased.
As demonstrated in the inset of Fig.\ \ref{fig:twa_precision} the deviations from the exact result are compatible with a power law scaling; $N^{-2}$ is shown as orientation.

%

\section{Semiclassical echo dynamics}
\label{section:echoes}
For our purposes we choose the perturbation Hamiltonian
$\hat H_p=\sum_\alpha J_\alpha \big(\hat c_\alpha^\dagger\hat c_{\alpha+1}+h.c.\big)$
with normally distributed random couplings $J_\alpha$ (variance $J^2=\overline{J_\alpha^2}$) and corresponding Weyl symbol $\mathcal H_p=2\sum_\alpha J_\alpha\rho_{\alpha,\alpha+1}$.
Note that the dynamics under this Hamiltonian is captured exactly by the TWA, because it is quadratic.

\subsection{Echoes in finite systems}
\begin{figure}[t]
\includegraphics[width=.95\columnwidth]{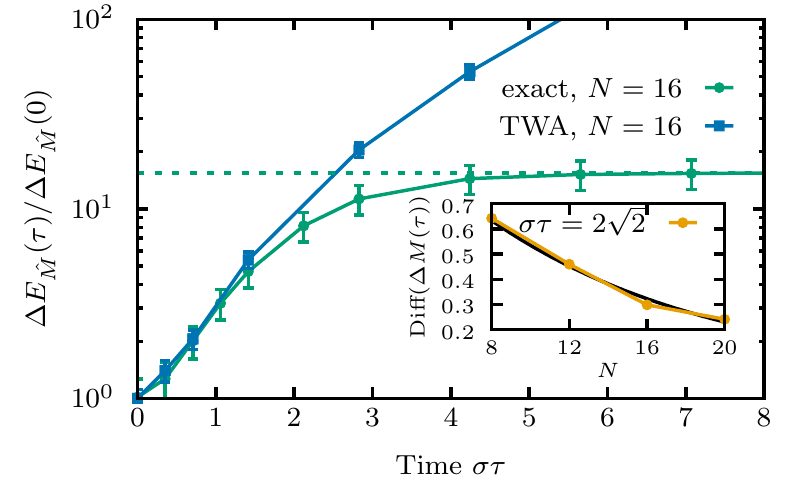}
\caption{Echo dynamics computed with TWA in comparison with exact results for $J\delta t=0.1$ and $N=16$ at quarter filling. The dashed line indicates the corresponding persistent echo peak height derived in Appendix \ref{app:finite_size}. The inset demonstrates that the normalized difference at the echo time is reduced as the system size is increased; the black line corresponds to an exponential fit.}
\label{fig:cmp_echo}
\end{figure}

In Fig.\ \ref{fig:cmp_echo} we compare the result for $\Delta E_M(\tau)$ given in Eq.\ \eqref{eq:echo_divergence} obtained from TWA with the exact dynamics. 
The presented data includes a disorder average over $80$ realizations of both the SYK and the perturbation Hamiltonian.
In the initial uncorrelated state one quarter of the sites is filled and the rest is empty. 
We find with both methods that the echo deviates increasingly from the initial value as the waiting time $\tau$ is increased and the results are in good agreement at short times.
At long times, however, there is a clear discrepancy. 
In the result obtained from TWA the echo signal ultimately vanishes completely, meaning that $\Delta M(\tau\to\infty)=3/4$.
%
By contrast, the exact result saturates much earlier.
The reason for this is that for finite $N$ the overlap $\braket{\psi(\tau)|\hat P_{\delta t}|\psi(\tau)}$ is non-zero, resulting in an ever-persisting revival at the echo time \cite{Schmitt2017}.
The corresponding saturation value can be determined in the exact simulation and it is indicated in Fig.\ \ref{fig:cmp_echo} by the dashed line; see Appendix \ref{app:finite_size} for details.
This persisting echo, however, vanishes for $N\to\infty$, because, typically, the Loschmidt echo is exponentially suppressed by the system size, $|\braket{\psi(\tau)|\hat P_{\delta t}|\psi(\tau)}|^2\sim e^{-Nr(\delta t)}$ with an intensive rate function $r(t)$. Hence, the limits $N\to\infty$ and $\delta t\to0$ do not commute.
Correspondingly, the normalized difference between exact and TWA data at the echo time, Diff($\Delta E_{\hat M}(\tau)$), is reduced when the system size is increased, as indicated in the inset of Fig.\ \ref{fig:cmp_echo}.
In Appendix \ref{app:finite_size} we include further data supporting the anticipated vanishing of the persistent echo in the exact dynamics for $N\to\infty$.
This disappearing of an intrinsic difference between TWA and exact echo dynamics goes along with a generally improved accuracy of the TWA as discussed above. 
Therefore, we expect that in the limit $N\to\infty$ results from TWA and exact dynamics will converge.
Since we find in addition that the TWA result for the echo dynamics is independent of $N$ (see Appendix \ref{app:finite_size}), we conclude that the TWA results obtained for large but finite $N$ constitute a good approximation of the behavior in the large $N$ limit. 

With our resources for the exact dynamics, however, $N=20$ is the largest value we can reach due to the large number of nonvanishing matrix elements in the SYK Hamiltonian and the disorder average necessary to perform a meaningful finite size analysis.
For these finite systems the persisting echo can be considered to be a genuine quantum characteristic.
The TWA, applicable in the semiclassical limit, does not capture this feature, because its origin is the non-vanishing overlap between the quantum states before and after application of the perturbation operator in combination with the unitarity of quantum time evolution.

\begin{figure}[t!]
\includegraphics[width=.95\columnwidth]{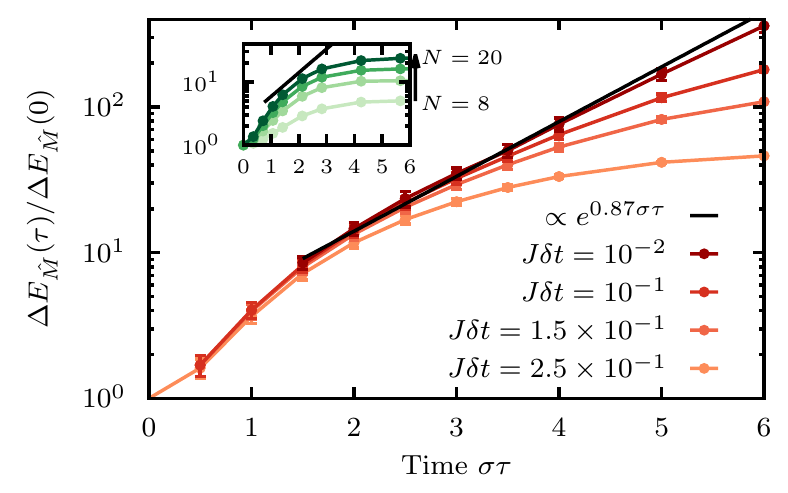}
\caption{TWA results for the divergence from the perfect echo computed for $N=20$ modes. As the perturbation strength $\delta t$ is decreased the regime of exponential growth is extended, allowing for the identification of a Lyapunov exponent. The inset shows exact results for system sizes $N=8,12,16,20$ for $J\delta t=0.1$. These exact data are compatible with convergence towards the TWA result as $N\to\infty$.}
\label{fig:echo_decay}
\end{figure}
\begin{figure*}[ht]
\center
\includegraphics[width=\textwidth]{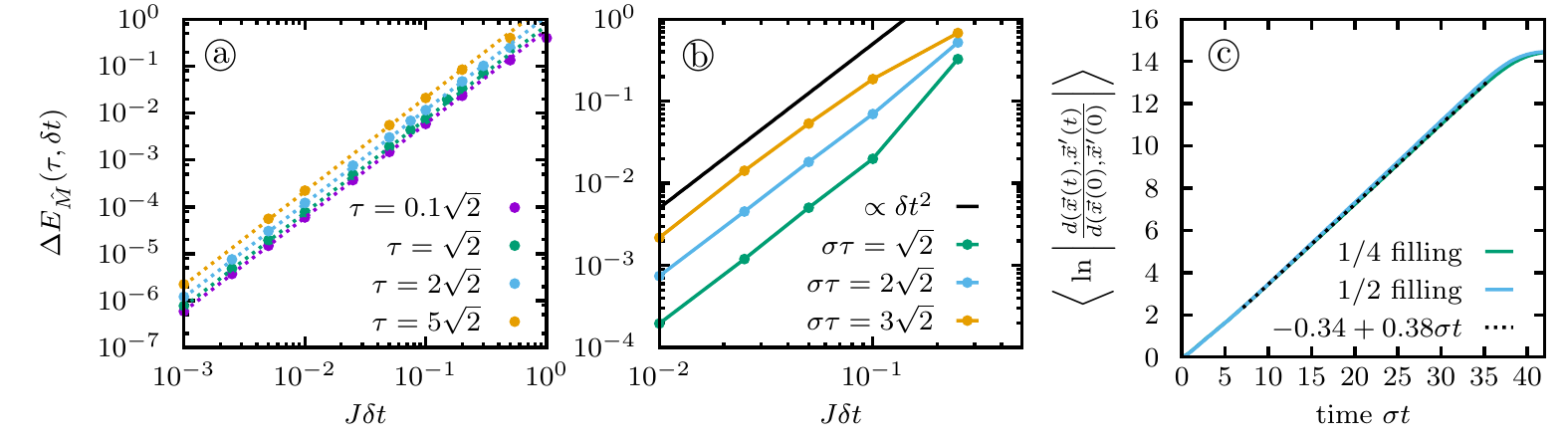}
\caption{(a) Echo divergence $\Delta M(\tau)$ for different fixed $\tau$ as function of the perturbation strength $\delta t$. Comparison of data obtained from exact full echo dynamics (dots) with the quadratic term in Eq. \eqref{eq:echo_divergence} alone (dashed lines). The double commutator is the single contribution to the echo divergence over a large range of $\tau$ and $\delta t$. (b) Corresponding TWA result. (c) Averaged divergence of initially close-by trajectories in phase space under TWA dynamics. A linear fit yields the estimate for the classical Lyapunov exponent.}
\label{fig:shorttime}

\end{figure*}
\subsection{Signature of a butterfly effect in echo dynamics}
In Fig.\ \ref{fig:echo_decay} we show TWA results for the divergence from the perfect echo as defined in Eq.\ \eqref{eq:echo_divergence}.
%
%
%
%
After the short time period the data exhibit a clear exponential growth of the difference to the perfect echo although the observable is bounded.
We find that the parameter $\delta t$ that determines the smallness of the perturbation controls the extent of the regime, where the exponential law is observed.
In direct analogy to classical chaos the exponential divergence of the perturbed echo from the perfect echo allows to identify a Lyapunov exponent $\lambda_L$.
A fit to the data in Fig.\ \ref{fig:echo_decay} yields $\lambda_L\approx0.87\sigma$, which is in good quantitative agreement with a result for the Lyapunov exponent in the limit of high temperature obtained via a diagrammatic large-$N$ expansion and exact numerics.\cite{Kobrin2018,Roberts2018} Note that our convention for the coupling constants differs from Ref. \cite{Roberts2018} by a factor $\sqrt{2}$.
In Appendix \ref{app:corr_echo} we include results for another observable, namely density-density correlations, showing exponential divergence with the same rate.
In the following we will discuss the origin of this exponential divergence in more detail.

\subsection{Role of the double commutator}
In the exact echo dynamics we observe that the quadratic term of Eq.\ \eqref{eq:echo_divergence}, in fact, is the only relevant contribution for a large range of perturbation strengths and irrespective of the waiting time.
Fig. \ref{fig:shorttime}a shows exact data for $\Delta E_{\hat M}(\tau)$ in comparison with the quadratic term $\frac12\braket{\psi_0|[\hat H_p(\tau),[\hat H_p(\tau),\hat M]]|\psi_0}\delta t^2$ as function of the perturbation strength $\delta t$ for different waiting times $\tau$.
Both coincide perfectly for $J\delta t<0.5$.

Even though the TWA does not capture the persistent echo, Fig.\  \ref{fig:shorttime}b demonstrates that the semiclassical echo dynamics exhibit the same quadratic dependence on the perturbation strength $\delta t$ in the regime of exponential growth.
Deviations from the quadratic scaling only occur when $\Delta E_{\hat M}(\tau)$ begins to saturate.
This supports the assertion that in Eq. \eqref{eq:echo_divergence} the second order term is the single contribution responsible for the exponential sensitivity to the imperfection in the time reversal protocol.

Similar to the OTOC \eqref{eq:otoc}, which is related to the square of the commutator of both operators, $|[V,W(t)]|^2$, 
expanding the double commutator reveals an out-of-time-order structure:
\begin{align}
	[\hat H_p(\tau),[\hat H_p(\tau),\hat M]]
	&=\hat H_p(\tau)^2\hat M+\hat M\hat H_p(\tau)^2
	\nonumber\\&\quad
	-2\hat H_p(\tau)\hat M\hat H_p(\tau)
	\label{eq:double_commutator}
\end{align}
In this expression the last term accounts for the butterfly effect.
For an extensive perturbation Hamiltonian $\hat H_p$ the double commutator becomes extensive at late times.
In the thermodynamic limit the double commutator can grow indefinetely such that it can govern the exponential divergence from the perfect echo irrespective of the higher order terms as long as $\braket{\psi_0|[\hat H_p(\tau),[\hat H_p(\tau),\hat M]]|\psi_0}\delta t^2\ll1$.
We deduce that only at late times higher order terms become important resulting in the approach to a constant.

The inference that the double commutator governs the exponential divergence in the echo dynamics is supported by the relation to the Lyapunov exponent of the classical TWA equations, which is discussed next.

\subsection{Classical Lyapunov exponent of the TWA equations}
The Lyapunov exponent occurring in the semiclassical echo dynamics can be related to the largest Lyapunov exponent of the dynamical system defined by the TWA equations of motion.
The largest classical Lyapunov exponent is defined as
\begin{align}
	\lambda_{\text{cl}}=\Big\langle\lim_{t\to\infty}\lim_{d(\vec x(0),\vec x'(0))\to0}\frac{1}{t}\ln\Big|\frac{d(\vec x(t),\vec x'(t))}{d(\vec x(0),\vec x'(0))}\Big|\Big\rangle
	\label{eq:class_lyapunov}
\end{align}
with coordinate vectors $\vec x(t)$ and $\vec x'(t)$ and $d(\vec x,\vec x')=\sqrt{\sum_{i}(x_i-x_i')^2}$ the Euclidian distance.
The time-dependence of the coordinate vectors is given by the equations of motion and $\braket{\cdot}$ indicates the classical average over an ensemble of trajectories.

To estimate the Lyapunov exponent of the TWA equations of motion we average the divergence of an ensemble of initially close-by trajectories on a fixed time interval; details are given in Appendix \ref{app:lyapunov}.
Fig.\ \ref{fig:shorttime}c displays the resulting average $\braket{\ln\big|d(\vec x(t),\vec x'(t))/d(\vec x(0),\vec x'(0))\big|}$, which we computed for half and quarter filling with $d_0=10^{-8}$.
We find a clear linear dependence on time and a fit yields the classical Lyapunov exponent $\lambda_{\text{cl}}\approx0.34$.
The result varies only weakly as the filling is changed.

This value of $\lambda_\text{cl}$ is slightly less than half of $\lambda_L$, which we extracted before from the echo dynamics.
In the following we will argue that a factor of two between both is to be expected. We attribute the slight discrepancy to the different orders of averaging and taking the logarithm, resulting in a slightly smaller classical Lyapunov exponent as reported in Ref.\ \cite{Rozenbaum2017}.

The Weyl symbol of the double commutator in Eq.\ \eqref{eq:double_commutator} can be written in the form
\begin{align}
	&\big([\hat H_p(\tau),[\hat H_p(\tau),\hat M]]\big)_W=
	\nonumber\\&
	A^i_j\frac{\partial x_i(t)}{\partial x_j(0)}+B^i_j\frac{\partial x_i(0)}{\partial x_j(t)}
	+C_{ij}^{kl}\frac{\partial x_k(t)}{\partial x_i(0)}\frac{\partial x_l(0)}{\partial x_j(t)}
	\label{eq:dc_weyl}
\end{align}
with $\vec x$ the vector of $\rho$ and $\tau$ coordinates of the TWA equations (cf. Appendix \ref{app:dc}).
The modulus of all derivatives occurring in this expression grows with the classical Lyapunov exponent.
However, the sums of the single derivatives in the first two terms will cancel, because they correspond to linear response.
Hence, if the terms in the quadratic contribution do not cancel, at late times
\begin{align}
	\big([\hat H_p(\tau),[\hat H_p(\tau),\hat M]]\big)_W\sim e^{2\lambda_\text{cl}t}\ .
\end{align}
The Weyl symbols of higher order commutators would contain growth rates that are higher multiples of $\lambda_\text{cl}$.
Since we only observe the factor of two in the echo dynamics, we conclude that the quadratic term in Eq.\ \eqref{eq:echo_divergence} is in fact the one that is relevant for the butterfly effect.

\section{On the importance of including quantum fluctuations}
\label{sec:TWAvsMF}
It is worthwhile elaborating more on the importance of quantum fluctuations for the dynamics of the SYK model in the semiclassical limit. In the following we will contrast mean field dynamics, which captures only fluctuations on the Gaussian level of single Fermion operators, against dynamics, which includes fluctuations that are Gaussian on the level of Fermionic bilinears.

The TWA equations of motion presented in Section \ref{subsec:eom} are essentially mean field equations of motion. In the following we aim to outline the key difference between TWA and the mean field approximation, namely the fact that TWA captures fluctuations which are essential for the dynamics of the SYK model. The importance of fluctuations is due to the fact that the microscopic degrees of freedom are fermions, which are always strongly fluctuating. This is, for example, in contrast to the semiclassical limit of large spins, where the fluctuations vanish in the limit of large spin.

The most general equations of motion for a mean field approximation are
\begin{widetext}
\begin{align}
	\im\frac{d}{dt}\rho_{\alpha\beta}&=
	\frac{2}{N^{3/2}}\sum_{ijkl}J_{ijkl}\delta_{\alpha l}\Big(
	\tau_{ji}^*\tau_{\beta k}+\delta_{ik}\rho_{j\beta}+2\rho_{j\beta}\rho_{ik}
	\Big)
	-(\alpha\leftrightarrow\beta)^*
	\nonumber\\
	\im\frac{d}{dt}\tau_{\alpha\beta}
	&=\frac{2}{N^{3/2}}\sum_{ijkl}J_{ijkl}\delta_{j\alpha}
	\Big(\delta_{il}\tau_{k\beta}-\rho_{i\beta}\tau_{kl}-2\rho_{il}\tau_{\beta k}\Big)
	-(\alpha\leftrightarrow\beta)
	\label{eq:mean_field_eom}
\end{align}
\end{widetext}
These equations are obtained under the assumption that the quantum state remains Gaussian for all times. In that case a Wick theorem can be used to split all higher order correlations into products of two-point functions, which correspond to the resulting phase space variables.

The mean field Hamiltonian corresponding to Eq.\ \eqref{eq:mean_field_eom}  encorporates all possibilities, meaning that there are classical fields coupling to pairing terms as well as hopping and local potentials. It turns out (see Fig.\ \ref{fig:red_eom} below) that to approximate the SYK dynamics it is sufficient to consider a much simpler mean field Hamiltonian including only pairing operators, given that the quantum fluctuations in the initial state are taken into account. This simpler mean field Hamiltonian takes the form
\begin{align}
	\hat H=\frac{1}{\sqrt{2N}}\sum_{ij}(\Delta_{ij}(t)\hat c_i^\dagger\hat c_j^\dagger+h.c.)
\end{align}
where the classical field
\begin{align}
	\Delta_{ij}(t)=\frac{1}{2N}\sum_{kl}J_{ijkl}\braket{\hat c_k\hat c_l}_t=\frac{1}{2N}\sum_{kl}J_{ijkl}\tau_{kl}(t)
\end{align}
is determined self-consistently. The resulting equations of motion constitute a reduction of Eq.\ \eqref{eq:twa_eom}:
\begin{align}
\im\frac{d\rho_{\alpha\beta}}{dt}&=-\frac{2}{\sqrt{2N}}\sum_k\Delta_{k\alpha}(t)^*\tau_{\beta k}-(\alpha\leftrightarrow\beta)^*
	\nonumber\\
	\im\frac{d\tau_{\alpha\beta}}{dt}&=\frac{2}{\sqrt{2N}}\sum_j\Delta_{\alpha j}(t)\rho_{j\beta}-(\alpha\leftrightarrow\beta)
	\label{eq:rand_sc_eom}
\end{align}

In the mean field approximation the initial condition of the phase space variables is fixed by the expectation values in the initial state,
\begin{align}
	\rho_{\alpha\beta}(0)&=\braket{\hat c_\alpha^\dagger\hat c_\beta}_{t=0}-\frac{\delta_{\alpha\beta}}{2}\nonumber\\
	\tau_{\alpha\beta}(0)&=\braket{\hat c_\alpha\hat c_\beta}_{t=0}=0
\end{align}
This means, however, that mean field dynamics with Eq.\ \eqref{eq:rand_sc_eom} is trivial, because $\braket{\hat c_i\hat c_j}_{t=0}=0$; in the mean field approximation the system remains stationary at all times. Non-trivial dynamics is only initiated by fluctuations of the fermionic bilinears in the initial state. These fluctuations can be included by stochastic sampling of the initial condition as we will discuss below.

\begin{figure}[t]
\includegraphics[width=.9\columnwidth]{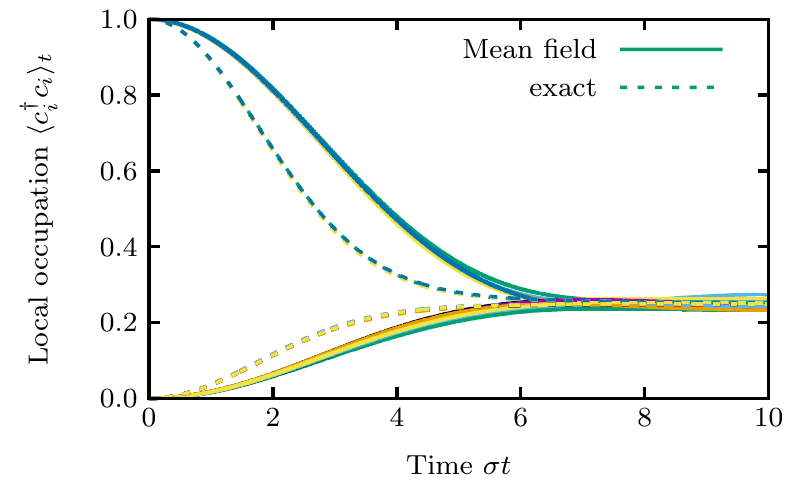}
\caption{Expansion dynamics from the uncorrelated initial stated as observed in the single mode occupation numbers $n_i(t)=\braket{\hat c_i^\dagger\hat c_i}_t$. The solid lines correspond to mean field dynamics based on the most general equations of motion, Eq.\ \eqref{eq:mean_field_eom}. The dashed lines were obtained by computing the full quantum dynamics. The data shown are for $N=20$ at quarter filling.}
\label{fig:mf}
\end{figure}

Within the mean field approximation non-trivial dynamics is obtained when considering the equations of motion given in Eq.\ \eqref{eq:mean_field_eom}. These equations account for Gaussian fluctuations on the level of single fermion operators. Fig.\ \ref{fig:mf} shows a result for relaxation dynamics obtained in the mean field approximation using Eq.\ \eqref{eq:mean_field_eom} starting with an uncorrelated state with occupation imbalance just as in the main text. Although the general shape of the decay is captured quite well, the decay time scale differs from the corresponding exact result. In the mean field approximation the relaxation turns out to be too slow. This discrepancy between mean field dynamics and exact dynamics was already observed in Ref.\ \cite{Davidson2017}, where, however, different mean field approximations were considered.

\begin{figure}[t]
\includegraphics[width=.9\columnwidth]{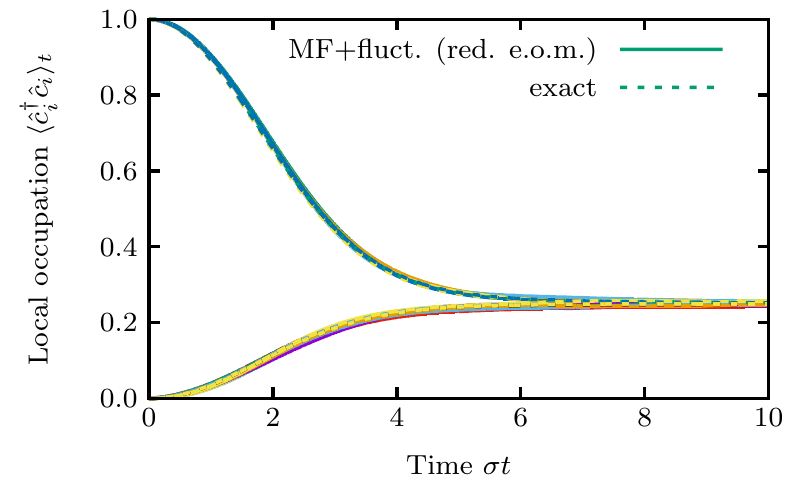}
\caption{Expansion dynamics from the uncorrelated initial stated as observed in the single mode occupation numbers $n_i(t)=\braket{\hat c_i^\dagger\hat c_i}_t$. The solid lines correspond to dynamics obtained based on the reduced mean field equations of motion, Eq.\ \eqref{eq:rand_sc_eom}, including fluctuations in the initial state by stochastic sampling of the initial conditions. The dashed lines were obtained by computing the full quantum dynamics. The data shown are for $N=20$ at quarter filling.}
\label{fig:red_eom}
\end{figure}

In order to accurately describe the relaxation dynamics it is essential to capture fluctuations of the fermionic bilinears correctly. This can be achieved by including Gaussian fluctuations of the phase space variables\footnote{Note that Gaussian fluctuations on the level of the phase space variables (fermionic bilinears) includes the connected parts of the fermionic four point function.} by stochastic sampling and an averaging of the resulting trajectories. This approach is essentially equivalent to stochastic sampling from the Wigner function of the initial state as it is done in the TWA.  Fig.\ \ref{fig:red_eom} displays the result for relaxation dynamics obtained in this approximation using the equations of motion of the simple mean field Hamiltonian, Eq.\ \eqref{eq:rand_sc_eom}, supplemented with fluctuations of the initial conditions. The comparison with the exact result shows very good agreement. Hence, we conclude that the relaxation dynamics is mainly driven by two-particle fluctuations, which are included in the TWA, but not in the mean field approximation.

A similar approach to incorporate quantum fluctuations in phase space dynamics has already been introduced in Ref. \cite{Damle1996}. However, as it is evident from the discussion above, there are various ambiguities, for which there is no a-priori resolution. Nevertheless, the corresponding choices might affect the resulting physical quantities. For example, the additional terms occurring in Eq.\ \eqref{eq:mean_field_eom}, which are irrelevant for the dynamics in our case, might be important under different circumstances. \cite{Polkovnikov2010} The TWA provides a consistent mathematical framework to set up the equations of motion and to incorporate fluctuations. The remaining ambiguity in choosing the bilinears based on which the phase space is constructed corresponds to finding the decoupling scheme where the saddle point approximation becomes asymptotically exact (cf. Section \ref{sec:onTWA}).

As a final remark we would like to mention that the shortcomings of the mean field approximation are also reflected in the fact that with mean field only a sub-extensive part of the spectrum can be captured\cite{Scaffidi2017} and only fluctuations as included in TWA render the energies extensive.

\section{Discussion}\label{sec:discussion}
We found that the exponential divergence from the perfect echo in the semiclassical dynamics is due to the growth of an out-of-time-order double commutator of the form $[\hat V(\tau),[\hat V(\tau),\hat W(0)]]$.
This assertion is based on the small perturbation expansion in Eq.\ \eqref{eq:echo_divergence}, which does not rely on any semiclassical approximation.
In future work the structure and characteristic behavior of these objects should be further explored, in particular with regard to the sensitivity of genuine quantum dynamics far from a classical limit to small perturbations.

Regarding irreversibility our result implies that the dynamics of the SYK model is irreversible in the same sense as a chaotic classical system: Any imperfection in the time reversal procedure leads to an exponential divergence from the perfect echo and substantial improvement is prohibitively expensive, because the Lyapunov exponent is perturbation-independent.

\begin{acknowledgments}
The authors acknowledge helpful discussions with S. Davidson.
This work was supported through SFB 1073 (project B03) of the
Deutsche Forschungsgemeinschaft (DFG).
M.S.\ acknowledges support by the Studienstiftung des Deutschen Volkes and through the Leopoldina Fellowship Programme of the German National Academy of Sciences Leopoldina (LPDS 2018-07) with additional support from the Simons Foundation.
D.S.\ acknowledges support from the FWO as post-doctoral fellow of the Research Foundation – Flanders and CMTV.
A.P.\ acknowledges support by NSFGrants No. DMR-1506340, No. DMR-1813499 and AFOSR Grand No. FA9550-16-1-0334.
For the numerical computations Armadillo \cite{armadillo} and ITensor \cite{itensor} were used.
\end{acknowledgments}

\appendix

\begin{figure}[b]
\includegraphics{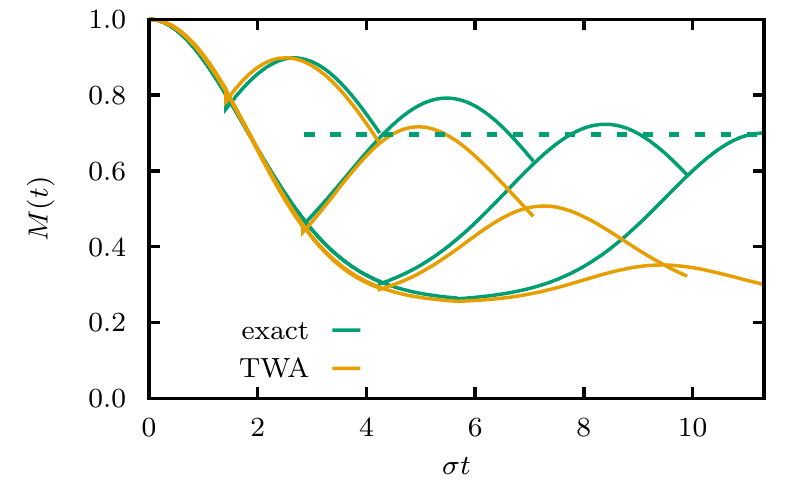}
\caption{Full time evolution under imperfect effective time reversal as obtained by TWA in comparison with exact dynamics for different forward times $\tau$ with system size $N=20$ at quarter filling and $J\delta t=0.25$. The exact dynamics show a persistent echo signal, whereas the TWA echo vanishes at long forward times. The dashed line indicates the persistent peak height as given by Eq.\ \eqref{eq:persistent_echo_height}.}
\label{fig:full_echo_dynamics}
\end{figure}
\section{Finite size analysis}
\label{app:finite_size}
For finite mode number $N$ the perturbed state will always have a nonvanishing overlap with the unperturbed state, $|\braket{\psi(\tau)|\hat P_{\delta t}|\psi(\tau)}|>0$.
Accordingly, we can decompose $\hat P_{\delta t}\ket{\psi(\tau)}=\cos(\alpha_{\delta t})\ket{\psi(\tau)}+\sin(\alpha_{\delta t})\ket{\phi}$ by introducing the ``orthogonal component" $\ket{\phi}$ with $\braket{\psi(\tau)|\phi}=0$.
Considering this decomposition it becomes evident that the remaining ``parallel component" of the perturbed state leads to an ever persisting echo at time $t=2\tau$:
\begin{align}
	&\braket{\psi(\tau)|\hat P_{\delta t}^\dagger e^{-\im\hat H\tau}\hat Me^{\im\hat H\tau}\hat P_{\delta t}|\psi(\tau)}\nonumber\\
	&=\cos^2(\alpha_{\delta t})\braket{\psi_0|\hat M|\psi_0}+\sin^2(\alpha_{\delta t})\braket{\phi|e^{-\im\hat H\tau}\hat Me^{\im\hat H\tau}|\phi}
	\nonumber\\&\quad
	+\sin(2\alpha_{\delta t})\text{Re}\big(\braket{\psi_0|\hat Me^{\im\hat H\tau}|\phi}\big)
\end{align}
For finite $N$ there is a time-independent contribution proportional to the initial value of the observable, $\braket{\psi_0|\hat M|\psi_0}$, and the overlap of the perturbed and unperturbed state, $\cos^2(\alpha_{\delta t})=|\braket{\psi(\tau)|\hat P_{\delta t}|\psi(\tau)}|^2$.
At late times the expectation value in the second term will attain an equilibrium value $M_\phi^\infty=\lim_{\tau\to\infty}\braket{\phi|e^{-\im\hat H\tau}\hat Me^{\im\hat H\tau}|\phi}$ and the overlap in the third term will vanish. Therefore, the persistent echo peak height at large $\tau$ is given by
\begin{align}
	\lim_{\tau\to\infty}E_{\hat M}(\tau)=\cos^2(\alpha_{\delta t})\braket{\psi_0|\hat M|\psi_0}+\sin^2(\alpha_{\delta t})M_\phi^\infty\ .
	\label{eq:persistent_echo_height}
\end{align}
Exemplary results for the dynamics including effective time reversal are shown in Fig.\ \ref{fig:full_echo_dynamics}.
In the thermodynamic limit, $N\to\infty$, we will have $\alpha_{\delta t}=\pi/2$, i.e. the contribution given by the initial expectation value of $\hat M$ vanishes and we obtain
\begin{align}
	\lim_{N\to\infty}\lim_{\tau\to\infty}E_{\hat M}(\tau)=M_\phi^\infty\ .
\end{align}

\begin{figure}[t]
\includegraphics{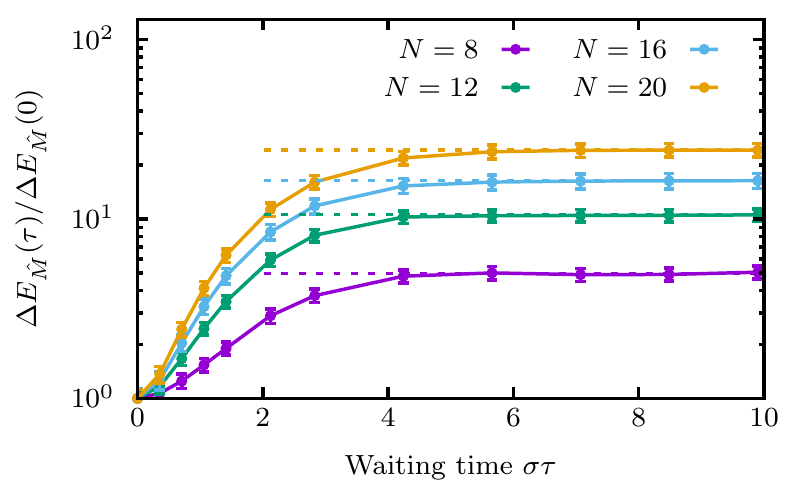}
\caption{Finite size analysis of exact results for the echo dynamics. The dashed lines indicate the saturation values obtained from the overlap of the perturbed with the unperturbed state. Here, $J\delta t=0.1$.}
\label{fig:ed_finite_size}
\end{figure}
Moreover, the window for possible exponential divergence from the perfect echo has a fixed size for a given finite $N$. This window cannot be increased by reducing $\delta t$, which is evident from Eq.\ \eqref{eq:echo_divergence} of the main text. In a finite system the expectation value of the double commutator is bounded for all times $\tau$, $|\braket{\psi_0|[\hat H_p(\tau),[\hat H_p(\tau),\hat O]]|\psi_0}|<C(N)$. Therefore, in the limit of small $\delta t$
\begin{align}
	1\leq\Big|\frac{\Delta E_{\hat O}(\tau)}{\Delta E_{\hat O}(0)}\Big|<\frac{C(N)}{|\braket{\psi_0|[\hat H_p,[\hat H_p,\hat O]]|\psi_0}|}\ .
\end{align}

In the following we present data for the variation of the echo signal $E_{\hat M}(\tau)$ with changing system sizes, which supports our assertion that the persistent echo vanishes in exact quantum dynamics. 
For a faithful investigation of finite size effects disorder averaging is essential, because fluctuations introduced by adding new randomly coupled degrees of freedom can otherwise spoil the analysis.

In Fig.\ \ref{fig:ed_finite_size} we show exact results for the divergence from the perfect echo for different system sizes, including a disorder average over 80 realizations. 
The dashed lines indicate the saturation value of the persistent echo computed directly according to Eq.\ \eqref{eq:persistent_echo_height}, where at quarter filling $M_\phi^\infty=1/4$.
We find very good agreement of the echo at late times with this value.
As discussed in the main text and earlier in this section the saturation value increases as the system size is increased. This corresponds to the vanishing of the persistent echo in the thermodynamic limit.

\begin{figure}[t]
\includegraphics{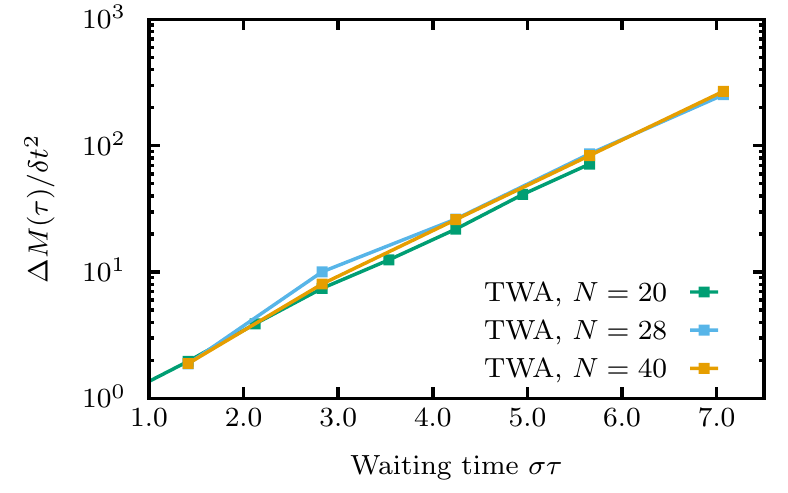}
\caption{Finite size analysis of results for the echo dynamics obtained using TWA.}
\label{fig:twa_finite_size}
\end{figure}

Fig.\ \ref{fig:twa_finite_size} displays TWA results for the divergence from the perfect echo for different system sizes. In this case we find that the results are almost identical despite a doubling of the system size.

Combining both results with the expectation that TWA becomes exact in the thermodynamic limit we conclude that the TWA result gives already at finite system sizes a good approximation of the result in the thermodynamic limit and with increasing $N$ the exact results will converge to this.

\begin{figure}[b]
\includegraphics{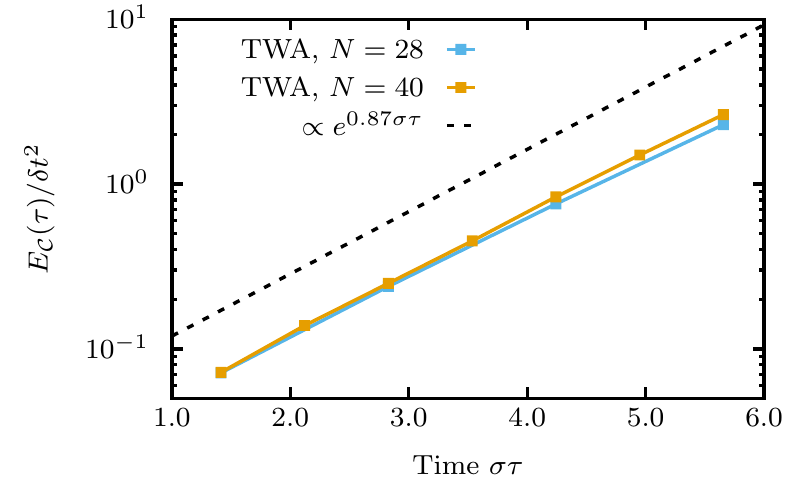}
\caption{Echo observed in the correlation average $\mathcal C$ (cf. Eq.\ \eqref{eq:corr_avg}) as a function of waiting time $\tau$. The exponential rate is the same as in the case of the occupation imbalance.}
\label{fig:corr_echo}
\end{figure}

\section{Echo in density-density correlation}
\label{app:corr_echo}
In addition to the occupation imbalance $\hat M$ presented in the main text we investigated echos in density-density correlations. We consider the average correlation
\begin{align}
	\mathcal C(t)=\frac{2}{N(N-1)}\sum_{i<j}\big|\braket{\hat n_i\hat n_j}_t-\braket{\hat n_i}_t\braket{\hat n_j}_t\big|
	\label{eq:corr_avg}
\end{align}
with $\hat n_i=\hat c_i^\dagger\hat c_i$.

Fig.\ \ref{fig:corr_echo} shows the dynamics of the echo $E_{\mathcal C}(\tau)$ as defined in Eq.\ \eqref{eq:echo_definition}. 
With increasing waiting time $\tau$ we find also for the correlation average $\mathcal C$ an exponential divergence from the perfect echo. The exponential rate is the same as in the case of the occupation imbalance.

\section{Approach to determine the classical Lyapunov exponent}
\label{app:lyapunov}
A common numerical method to determine the largest classical Lyapunov exponent
\begin{align} 
	\lambda_{\text{cl}}=\Big\langle\lim_{t\to\infty}\lim_{d(\vec x(0),\vec x'(0))\to0}\frac{1}{t}\ln\Big|\frac{d(\vec x(t),\vec x'(t))}{d(\vec x(0),\vec x'(0))}\Big|\Big\rangle
	\label{eq:class_lyapunov}
\end{align}
is to integrate the equations of motion of two close-by initial conditions $\vec x(0)$ and $\vec x'(0)$ with a small fixed $d(\vec x(0),\vec x'(0))=d_0$ and evaluate the ratio $d(\vec x(t),\vec x'(t))/d_0$ at a fixed time $t$.
Then $\vec x'$ is reinitialized such that $d(\vec x(t),\vec x'(t))=d_0$ and the equations of motion are integrated for another interval $t$, before the ratio of initial and final distances is evaluated again.
This procedure is iterated and the samples of $t^{-1}\ln\big|d(\vec x(t),\vec x'(t))/d(\vec x(0),\vec x'(0))\big|$ are averaged to obtain an estimate of the classical Lyapunov exponent \eqref{eq:class_lyapunov}.

To estimate the Lyapunov exponent of the TWA equations of motion we employed a similar approach.
In this case $\vec x\equiv(\rho_{\alpha,\beta},\tau_{\alpha,\beta})$.
During a sequence of integration and reinitialization in turns we average $\ln\big|d(\vec x(t),\vec x'(t))/d(\vec x(0),\vec x'(0))\big|$ on the whole interval $0<t<t_\text{max}$.
Additionally, we average over many such sequences with initial conditions drawn from the Wigner function of the initial state under consideration.
In this way we obtained the result shown in Fig.\ \ref{fig:shorttime}c in the main text.

\section{Structure of the Weyl symbol of the double commutator}
\label{app:dc}
Let us denote the set of phase space variables by $\vec x$. In our case both $H_p$ and $M$ are linear in TWA variables, which means that the Bopp operators take the form
\begin{align}
	\vec H_p=h(\vec x)+\sum_i h_i(\vec x)\frac{\partial}{\partial x_i}
\end{align}
and
\begin{align}
	\vec M=m(\vec x)+\sum_i m_i(\vec x)\frac{\partial}{\partial x_i}
\end{align}
where $h(\vec x), h_i(\vec x), m(\vec x), m_i(\vec x)$ are some functions of the coordinates.

\begin{widetext}
Plugging this into the double commutator yields the Weyl symbol
\begin{align}
	&\big([\hat H_p(t),[\hat H_p(t),\hat M]]\big)_W\nonumber\\
	&=
	\Big(h(\vec x(t))+\sum_i h_i(\vec x(t))\frac{\partial}{\partial x_i(t)}\Big)\Big(h(\vec x(t))+\sum_j h_j(\vec x(t))\frac{\partial}{\partial x_j(t)}\Big)m(\vec x)
	\nonumber\\&\quad
	+\Big(m(\vec x)+\sum_i m_i(\vec x)\frac{\partial}{\partial x_i}\Big)
	\Big(h(\vec x(t))+\sum_j h_j(\vec x(t))\frac{\partial}{\partial x_j(t)}\Big)h(\vec x(t))
	\nonumber\\&\quad
	-2\Big(h(\vec x(t))+\sum_i h_i(\vec x(t))\frac{\partial}{\partial x_i(t)}\Big)
	\Big(m(\vec x)+\sum_j m_j(\vec x)\frac{\partial}{\partial x_j}\Big)h(\vec x(t))
	\nonumber\\
	&=\sum_{ij}h_i(\vec x(t))h_j(\vec x(t))\frac{\partial}{\partial x_i(t)}\frac{\partial}{\partial x_j(t)}m(\vec x)
	+\sum_{ij}h_i(\vec x(t))\frac{\partial h_j(\vec x(t))}{\partial x_i(t)}\frac{\partial}{\partial x_j(t)}m(\vec x)
	\nonumber\\&\quad
	-2\sum_{ij}h_i(\vec x(t))\frac{\partial m_j(\vec x)}{\partial x_i(t)}\frac{\partial h(\vec x(t))}{\partial x_j}
	+\sum_{ij} m_i(\vec x)\frac{\partial h(\vec x(t))}{\partial x_j(t)}\frac{\partial}{\partial x_i}h_j(\vec x(t))
	\nonumber\\&\quad
	-\sum_{ij}m_i(\vec x)h_j(\vec x(t))\frac{\partial}{\partial x_i}\frac{\partial h(\vec x(t))}{\partial x_j(t)}
\end{align}
Now we use the chain rule $\frac{\partial f(x_i(t_1))}{\partial x_j(t_2)}=\sum_k\frac{f(x_i(t_1))}{\partial x_k(t_1)}\frac{\partial x_k(t_1)}{\partial x_j(t_2)}$ wherever applicable, yielding
\begin{align}
	&\big([\hat H_p(t),[\hat H_p(t),\hat M]]\big)_W
	\nonumber\\
	&=
	\sum_{ijkl}\Big(h_i(\vec x(t))h_j(\vec x(t))\frac{\partial^2 m(\vec x)}{\partial x_k\partial x_l}\Big)\frac{\partial x_k}{\partial x_i(t)}\frac{\partial x_l}{\partial x_j(t)}
	+\sum_{jk}\Big(\sum_ih_i(\vec x(t))\frac{\partial h_j(\vec x(t))}{\partial x_i(t)}\frac{\partial m(\vec x)}{\partial x_k}\Big)\frac{\partial x_k}{\partial x_j(t)}
	\nonumber\\&\quad
	-2\sum_{ijkl}\Big(h_i(\vec x(t))\frac{\partial m_j(\vec x)}{\partial x_k}\frac{\partial h(\vec x(t))}{\partial x_l(t)}\Big)\frac{\partial x_k}{\partial x_i(t)}\frac{\partial x_l(t)}{\partial x_j}
	+\sum_{ik}\Big(\sum_j m_i(\vec x)\frac{\partial h(\vec x(t))}{\partial x_j(t)}\frac{\partial h_j(\vec x(t))}{\partial x_k(t)}\Big)\frac{\partial x_k(t)}{\partial x_i}
	\nonumber\\&\quad
	-\sum_{ik}\Big(\sum_jm_i(\vec x)h_j(\vec x(t))\frac{\partial^2 h(\vec x(t))}{\partial x_k(t)\partial x_j(t)}\Big)\frac{\partial x_k(t)}{\partial x_i}\ .
\end{align}
Since in our case $h(\vec x)$ and $m(\vec x)$ are linear in $\vec x$, the expression can be simplified to
\begin{align}
	&\big([\hat H_p(t),[\hat H_p(t),\hat M]]\big)_W
	\nonumber\\
	&=
	\sum_{jk}\Big(\sum_ih_i(\vec x(t))\frac{\partial h_j}{\partial x_i}\frac{\partial m}{\partial x_k}\Big)\frac{\partial x_k}{\partial x_j(t)}
	-2\sum_{ijkl}\Big(h_i(\vec x(t))\frac{\partial m_j}{\partial x_k}\frac{\partial h}{\partial x_l}\Big)\frac{\partial x_k}{\partial x_i(t)}\frac{\partial x_l(t)}{\partial x_j}
	+\sum_{ik}\Big(\sum_j m_i(\vec x)\frac{\partial h}{\partial x_j}\frac{\partial h_j}{\partial x_k}\Big)\frac{\partial x_k(t)}{\partial x_i}\ .
\end{align}
\end{widetext}
In this form the Weyl symbol corresponds to Eq.\ \eqref{eq:dc_weyl} in the main text.
This expression involves linear response type terms, which are linear in $\frac{\partial x_i(t)}{\partial x_j(0)}$, and terms that are quadratic in these derivatives. The linear terms should cancel such that they do not contribute to exponential growth; otherwise, also the response of the form $\{\hat H_p(\tau)^2,\hat M\}=\hat H_p(\tau)^2\hat M+\hat M\hat H_p(\tau)^2$ would grow exponentially.

\bibliography{refs}

\end{document}